# On Allocation Policies for Power and Performance

Dmytro Dyachuk* and Michele Mazzucco†‡
*University of Saskatchewan, Canada
†University of Tartu, Estonia
‡Software Technology and Applications Competence Centre (STACC), Estonia

*Abstract*—With the increasing popularity of Internet-based services and applications, power efficiency is becoming a major concern for data center operators, as high electricity consumption not only increases greenhouse gas emissions, but also increases the cost of running the server farm itself. In this paper we address the problem of maximizing the revenue of a service provider by means of dynamic allocation policies that run the minimum amount of servers necessary to meet user's requirements in terms of performance. The results of several experiments executed using Wikipedia traces are described, showing that the proposed schemes work well, even if the workload is non-stationary. Since any resource allocation policy requires the use of forecasting mechanisms, various schemes allowing compensating errors in the load forecasts are presented and evaluated.

## I. Introduction

A server farm is a collection of servers interconnected by high-speed, switched LANs that hosts content and runs applications (or services) accessed over the Internet. Data centers are an attractive alternative to enterprise systems because they offer economies of scale for network, power, cooling, administration, security and surge capacity [1]. However, because of the high users' expectations in terms of performance [2], [3], operating server farms in an energy efficient way is a challenging problem as service providers operate under stringent performance requirements, no matter whether they are dictated by users (see [4] and the references cited therein) or by Service Level Agreements (SLAs). At the moment, most of data center operators perform try to optimize their computational facilities for performance while ignoring energy consumption factors. However, given that the cost of servers purchase is comparable with the cost of the electricity required to run them for a 3 years period, power efficiency is becoming one of the major concerns for service providers, as it markedly affects the cost of running the data centers themselves. Consequently, it is of great importance to develop strategies aiming at reducing the power consumption while maintaining acceptable levels of performance.

While different techniques can be employed to reduce the server farm energy requirements (see Section V for more details), most of the efforts focus on single and specific architectural and technological improvements for achieving power reduction, ignoring the fact that power efficiency as a ratio of performance to power consumption is affected by both performance improvements and power reduction [5]. Hence, the only way to *significantly* reduce data centers' power consumption is to improve the server farm's utilization, *e.g.*, by tearing down servers in excess. The model we propose takes into account the fact that users are impatient and that servers energy consumption depends on servers' utilization. Therefore, we study and evaluate dynamic allocation policies aiming at maximizing the overall performance while minimizing the number of required servers when the traffic parameters are *not* known.

Next, we discuss how server farms can be designed and operated, and introduce the system model. Then we describe the model of user demand and service provision we propose, and present various energy aware allocation policies (Section III). Section IV contains several experiments we have carried out using the Wikipedia traces, while a survey of relevant related work is presented in Section V. Finally, we conclude the paper with future work and some remarks.

## II. The Model

Different models can be used to design a server farm, the most widely-used ones being *shared* and *dedicated* architectures. The former runs multiple applications on each server and multiplexes the server resources among these applications, while the latter does not share servers, but runs each application on a subset of the available servers, see Figure 1.

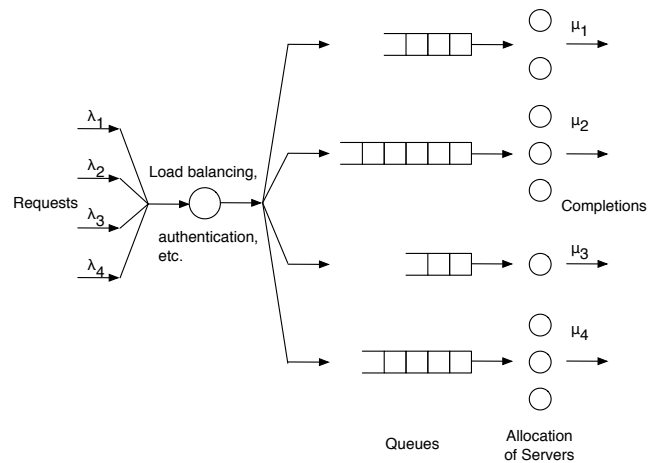

Fig. 1. Structure of a server farm designed according to the dedicated architecture.

Systems designed according to the dedicated architecture are usually deployed for running large clustered applications such as on-line mail services, web search engines or retail and brokerage services because the client workload of such applications makes resource sharing not feasible. In this paper we address the scenarios with dedicated server farms, where a web application is hosted on a set of physical servers. In this context, an immediate question that arises is how to manage the available resources in a cost-effective manner. Static resource allocation is attactive from its implementation prospective, however it entails excessive energy consumption as that policy attempts to satisfy the performance requirements under heavy traffic conditions via overprovisioning. Usually, the adoption of this policy results in poor utilization of resources which stays at 15-30%[1]. Thus, a natural solution to this problem would be to dynamically power servers up and down according to the load. Once a decision about how to partition the available servers has been made, it is possible to treat each subsystem (*i.e.*, service) in isolation of each other. Therefore, in the current work we tackle the problem of maximizing the revenue of a single service only.

It is assumed that $q$ identical servers are allocated to some service $i$. $h$ servers are switched on and are capable of handling the incoming requests (*jobs*, from now on), while the remaining $(q-h)$ are switched off. Powered down servers do not consume any electricity, but they can not be used for processing jobs.

Each server can process a maximum of $m$ jobs in parallel without marked interference. Such a limitation is inflicted by the number of available threads or processes. Such a behaviour can be modelled by assuming the presence of $m$ parallel servers on each physical machine or core, and thus a total of $S = qm$ servers are available, while $n = hm$ are running. If $n$ jobs are being processed (*i.e.*, all the threads on all servers are busy) newly arrived requests are placed in a queue (*i.e.*, on the load balancer site) whose size is assumed to be infinite. The queue acts according to the First-In-First-Out scheduling discipline, meaning that jobs are being served in the order of their arrivals. At job completion instants the server obtains another request from the queue, if any, otherwise it begins to idle. Idle servers still consume energy, but on a lower amount compared to that consumed by busy ones. Finally, none of the running servers is permitted to idle if there are requests waiting in the queue.

In order to model impatient customers (*e.g.*, clients that can click the Stop button in a browser while waiting for the server to respond) we assume that jobs sitting in the queue can time out. The exact time-outs associated with specific request are not known, however the distribution of the time-outs can be estimated at run-time. An abandoned request does not generate any profit, while each successfully processed job brings some profit. The source of the profit can be either advertisements or sales (in case of online merchants, such as Expedia). In the first case the estimation of the profit can be done directly as the agencies usually pay for an impression (display of a banner). In the second scenario not every request generates profit, however an average profit from serving a request can be computed as a ratio of the total profit over the number of served requests. For instance, 100,000 requests (page views) at the end bring 100$ of revenue from the sales. Thus, it can be said that on average each request brings 0.01 cents. Please note that the revenue model can be much more complex than those we have described. However, by employing transformations such as dividing gross income over the number of requests, one can easily calculate the average profit generated by each request.

One of the most important problems in applied queuing theory is the computation of the optimal number of servers needed in a multi-server queuing system. Given that running servers consume electricity and that user demand changes over the time, the provider must dynamically (*i.e.*, at run-time) decide how many servers to run in order to optimize his/her profit, *i.e.*, the optimal value of $n$. The extreme values, $n = 0$ and $n = S$, correspond to switch respectively off, or on, all available servers. We assume that changing the number of running servers does not affect service availability (this is certainly the case if data is widely replicated), but requires some time during which the servers being powered up/down consume energy without producing any revenue. Also, since lost jobs do not generate any revenue, the provider should ensure that the waiting time does not exceed users' patience, otherwise clients will start aborting their requests, see Figure 2.

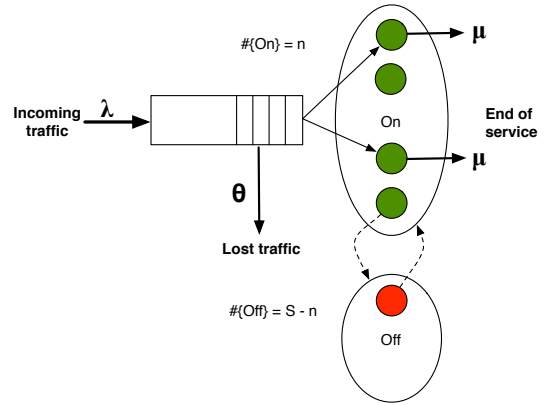

Fig. 2. System model. Jobs whose average job size is $1/\mu$ enter the system at rate $\lambda$ and abandon the system at rate $\theta$ while waiting.

During the intervals between consecutive policy invocations, the number of running servers remains constant. Those intervals, or 'observation windows', are used by the controlling software to collect the traffic statistics used by the allocation policy at the next decision epoch.

While different metrics can be used to measure the performance of a computing system, *e.g.*, average response time, throughput, or clients abandoning the system, here we are interested in maximizing the average revenue, $R$, earned by the service provider per unit time. That value can be estimated as

$$R = cT - rP, \qquad (1)$$

where $c$ is the income generated by each completed job, $T$ is the system's throughput, $r$ is the cost of electricity, and $P$ is the total average power consumed by the powered up servers.

[1] http://ganglia.wikimedia.org/

Although, we do not make any assumption about the relative magnitudes of charge and cost, the problem we present here is interesting mainly when they are close to each other. A charge higher than the provisioning cost would guarantee the provider a positive revenue (even thought not optimal) by switching on all servers, regardless of the load. On the other hand, if the charge is smaller than the cost, the provider would rather switch all servers off instead of serving incoming traffic.

## III. POLICIES

In this section we introduce the model of user demand and service provisioning used to dynamically control the number of operative servers.

We assume that jobs enter the system according to an independent Poisson process with rate $\lambda$, the service times are exponentially distributed with the mean of $1/\mu$, while $n$ servers accepting one job at a time are used to execute incoming user demand. In this paper we do not deal with SLAs; instead the performance requirements are dictated by users at runtime, see discussion in Section I. In other words we assume that a time-out policy is in operation: if a job entering the system does not acquire a server before its time-out period expires, the job is terminated and leaves the system without generating any revenue. While simple, this idea is rather powerful, since it lets us modelling HTTP time-outs as well as impatient customers. The latter are very important and should be taken into account when provisioning a distributed system, as [6] reports that 75% of people would not go back to a web site that took more than 4 seconds to load.

HTTP time-outs are in practice of known constant length, while users' patience is not. In order to make the problem analytically tractable it is assumed that both the user's patience and HTTP time-outs are i.i.d random variables distributed exponentially with mean $1/\theta$, with $\theta$ being referred to as the *abandonment rate*. Please note that the extreme values, $\theta = 0$ and $\theta = \infty$, correspond to jobs with no, or infinite patience. Also, we require the patience variables to be independent of all other model elements, namely arrival and service rates. Hence, we can treat the resulting model as an $M/M/n+M$ queue, also known as Erlang-A (where the 'A' stands for 'Abandonment') [7]. This model is not well known, but has a very nice property compared to the most popular Erlang-C queue [8] which can be employed for modelling users of an Internet service: no stability condition must be met, as jobs in excess are allowed to leave, and thus the queue never grows unbound. Therefore the load, $\rho = \lambda/\mu$, can exceed the number of operative servers as jobs abandonment reduces workload when the load is high. Hence, fewer servers are needed to guarantee the same level of performance under Erlang-A, compared to the traditional Erlang-C.

The transition diagram of this Markov process is illustrated in Figure 3. It is worth noting that while the instantaneous transition rate from state $j$ to state $(j+1)$ is equal to the arrival rate, the conditional departure rate from state $j$ to state $(j-1)$ at which jobs leave the system depends on the number of running servers as well as on the number of jobs present:

Case 1: $j \leq n$. The system behaves like an $M/M/\infty$ queue; all jobs in the system are being served without queueing, jobs leave the system at rate $\mu_j = j\mu$, and $(n-j)$ servers are idle.

Case 2: $j > n$. All servers are busy and $(j-n)$ jobs are queueing. The instantaneous completion rate does not depend on $j$ anymore, while the abandonment rate depends on the current number of jobs in the queue.

### A. Adaptive Policy

The 'Adaptive' heuristic we have introduced in [9] requires the solution of the steady state probability of the number of jobs present and, at each allocation decision epoch, considers the number of servers which are not powered down and the potential offered load. Also, it exploits the observation that as the size of the server farms grows, the system achieves economies of scale that makes it more robust against traffic variability. Hence, while violating the Markovian assumptions about the arrival, patience and service processes affects the average queue length, it does not substantially change the abandonment rate [10].

During system reconfigurations, the system is in a transient state which should be taken into account. Powering servers on/off takes on average $k$ units of time; hence, there are time and electricity losses because during state changes servers do not generate any profit, but do consume electricity. Also, system's reliability is affected by state changes, as hardware components tend to degrade faster with frequent power on/off cycles than with continuous operation. Therefore, each state change involves the following cost

$$Q = \frac{|\Delta n|}{t}(\sum_{i=1}^{l} d_i + kre_{max}), \qquad (2)$$

where $t$ is the length of the observation windows, $|\Delta n|$ is the number of servers that are switched on/off, $e_{max}$ is the power consumed per unit time during state changes, $k$ is the average time required to switch a server on/off, $d_i$ is the cost for a hardware component's state change, and $l$ is the number of components.

Hence, for a certain load $\rho$, the 'Adaptive' policy computes the expected revenue for consecutive values $n$ using a *binary search* algorithm and stops either when $R$ starts decreasing or, if that does not happen, when the increase becomes smaller than some value $\epsilon$ ($R$ is a concave function with respect to $n$):

$$\Delta r(n', n) = r(n') - r(n) - Q, \qquad (3)$$

where $r(n')$ is the expected revenue achieved when $n'$ servers are running and the load is $\rho$.

### B. QED Heuristic

The 'Adaptive' policy finds the 'optimal' number of servers to run (see the aforementioned remarks) using a binary search algorithm requiring $log(S)$ iterations. If one can provide a simpler heuristic that performs well, it may be worth accepting some slightly sub-optimal policy in order to avoid the search for the optimal one.

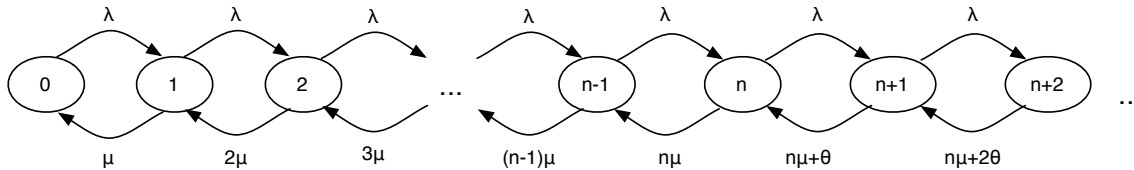

Fig. 3. State transition diagram.

Deciding on the number of servers to run requires to balance between the server farm's utilization and service quality (availability), with high utilization typically obtained at the cost of lower availability, and vice versa. Therefore, it is a common misconception that high utilization and good quality of service can not be achieved at the same time. When $\rho = n$, the system is critically loaded in the limit, and is said to be in the Quality and Efficiency-Driven (QED) regime, also known as Halfin-Whitt regime [11]. The behaviour of large server farms working in QED regime differs from that of Kingman's Law (i.e., delays/job losses are very common under heavy load) in that service quality is carefully balanced with server efficiency. Thus, we propose the following 'QED' heuristic. From the statistics collected during a window, estimate the arrival rate, $\lambda$, and average service time, $1/\mu$. For the duration of the next window, allocate the servers according to

$$n = \rho + \alpha\sqrt{\rho}, \qquad (4)$$

where the quantity $\alpha\sqrt{\rho}$ is used for dealing with stochastic variability, and where $\alpha$ represents the probability of all the servers being busy, e.g., when requests start queueing [12] (see discussion in Section IV-C).

Unfortunately, the arrival rate and consequently the load can not usually be estimated precisely. Hence, the QED policy should include an additional component which would let it compensate the uncertainty in respect to $\lambda$. Grassmann [12] suggests employing the following heuristic in order to deal with such kind of cases.

$$n = E(\rho) + \alpha\sqrt{E(\rho) + VAR(\rho)}. \qquad (5)$$

The term $VAR(\rho)$ is introduced in order to address the scenarios when the the information regarding the arrival rate is not deterministic, e.g., $\lambda$ is estimated using some forecasting techniques and $VAR(\rho)$ describes the variance of the prediction. Please note that $VAR(\rho) = VAR(\lambda)/\mu^2$, while $E(\rho)$ refers to the expected value of the load, i.e., $E(\rho) = E(\lambda)/\mu$ in our case.

## IV. EXPERIMENTS

### A. Arrival rate prediction

The policies we have described in Section III require an estimate of the arrival and service rates in order to decide the number of servers to run.

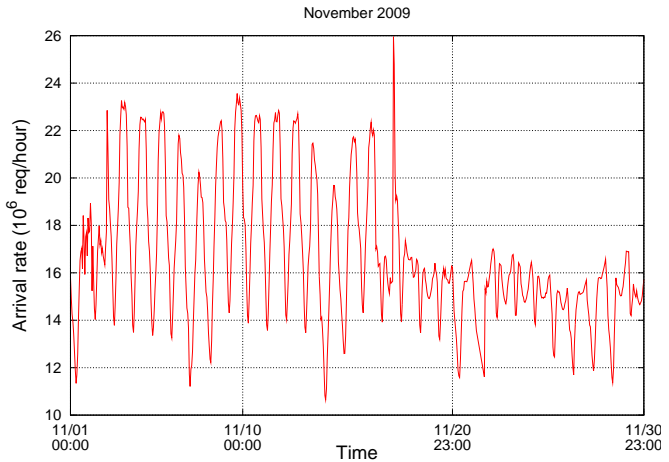

Fig. 4. Wikipedia workload for November, 2009.

As Figure 4 shows, the arrival rate in the real world scenarios is far from being stationary and fluctuations can reach the magnitude of several times. The fluctuations can have periodic components, which reflect the daily, weekly or even monthly interest in the service. However, in scenarios with large scale Internet services, e.g., Wikipedia, it is plausible to assume that the load is stationary for small time intervals (1 hour), i.e., variance and autocorrelation structure do not alter over time, or even if they do, such changes are relatively small. Thus, in such cases we suggest employing double exponential smoothing, which uses the smoothed value of the historical data and tries to predict the trend. Double exponential smoothing, also known as Winter's method, has a rather simple and straightforward implementation, and does not require significant amounts of historical data to be trained.

For any time period $t$, the smoothed value $S_t$ is found by solving the following equations

$$\begin{cases} S_t = \alpha\lambda_t + (1-\alpha)(S_{t-1} + b_{t-1}) \\ b_t = \gamma(S_t - S_{t-1}) + (1-\gamma)b_{t-1} \end{cases}, \qquad (6)$$

with $\alpha$ and $\gamma$ being two constants in the interval $0, \ldots, 1$, and $b_{t-1}$ being the trend from the previous period. In the above scheme, recent observations are given relatively more weight in forecasting than the older observations: the first equation adjusts the smoothed value $S_t$ adding $b_{t-1}$ to the last smoothed value, $S_{t-1}$, while the second equation is used to update the trend. Different schemes can be employed to initialize the algorithm; here we use

$$\begin{cases} S_1 = \lambda_1 \\ b_1 = \dfrac{\lambda_n - \lambda_1}{n - 1} \end{cases}, \quad (7)$$

while the best values of $\alpha$ and $\gamma$, i.e., those reducing the mean squared error, are computed using non-linear optimization techniques (we have used the Levenberg-Marquardt algorithm).

Having computed the smoothed and the trend values at time $t$, the forecast for the arrival rate at time $t+1$, $\lambda_{t+1}^F$, is computed as

$$\lambda_{t+1}^F = S_t + b_t. \quad (8)$$

Despite the fact the changes in the arrival rate are relatively small, a prediction with absolute precision is rather unlikely, and thus some error is inevitably introduced. The quality of the forecasting algorithm is evaluated based on how close the forecasting mechanism gets to the original value. Figure 5 contains a chart showing the distribution of the relative error of the forecasting produced by Winter's method when applied to Wikipedia traces of (November, 2009). The most interesting observation is that the distribution of the errors is symmetrical with the mean of zero. The histograms for the other months exhibit almost identical behavior, and thus they are not shown due to the limited space. The percentiles of the relative error are shown in Table I: as one can see, 90% of the arrivals rates can be estimated with 91% accuracy. Given the distribution of the relative error one can easily calculate the variance of the prediction, which later on will be used with QED heuristic, see Equation (5).

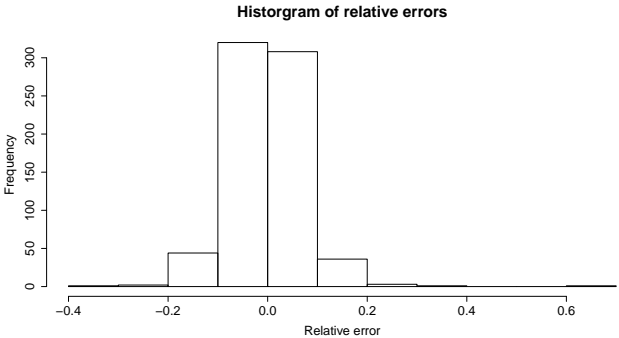

Fig. 5. Relative forecasting error for Winter's method applied to Wikipedia workload for November, 2009.

| Percentile | 90% | 95% | 99% |
|---|---|---|---|
| Error | 0.09 | 0.11 | 0.17 |

TABLE I
PERCENTILE OF THE RELATIVE ERROR.

In the following subsections we perform a set of experiments aiming at evaluating the efficiency of the proposed policies in term of revenue, energy consumption and jobs loss.

The experiments are conducted using a simulation model. For these experiments we assume that the data center has a Power Usage Effectiveness (PUE), a ratio of the total power over the power consumed by servers, of 1.7.

In order to reduce the experimentation space the following model parameters will have fixed values. Their values were chosen to reflect realistic scenarios with modern server farms.

- 250 physical servers equipped with 4 cores each i.e., $S = 1,000$.
- The power consumption of each four core machine ranges between 140 and 220 W [9]. In other words, each core (software server, see discussion in Section II) draws between 35 and 55 W. Since the server farm has a PUE factor of 1.7, the minimum and maximum power consumption are approximately $e_1 = 59$ and $e_2 = 94$ W per server.
- The cost for electricity, $r$, is 0.1 \$ per kWh[2].
- The average job size, $1/\mu$, is 0.1 seconds.
- The average user patience is 4 seconds, thus $\theta = 0.25$.
- Jobs are not strictly limited to CPU operations and thus a busy server requires 70% of the CPU, resulting in the power consumption of 69.58 Wh. Consequently, the electricity cost for each job is $2 \times 10^{-7}$\$ on average.
- Each successfully processed request generates a profit of $6.2 \times 10^{-6}$\$.

In each experiment server farm was exposed to the scaled version of the Wikipedia workload for November 2009 [Figure 4].

### B. Adaptive Policy

In the following subsections we present the behavior of the 'Adaptive' policy we have introduced in Section III-A. If the arrival rate is known this algorithm produces the optimal result; however this policy does not directly address the scenarios where the exact value of $\lambda$ is not known for the next configuration interval. In our previous work [9] we have demonstrated that this policy still exhibits satisfactory results in term of revenue even in the presence of some inaccuracies in the estimation of $\lambda$. Nonetheless, as shown in Figure 8 the error in the arrival rate estimation adversely affects the number of lost jobs. Therefore, in this work we introduce additional modifications to the adaptive search which allow optimizing the revenue while reducing the number of the lost jobs. More precisely, we suggest performing a *slight* overprovisioning in order to deal with unexpected traffic spikes. We try three different approaches of performing overprovisioning and evaluate their effect on the revenue, consumed energy and lost jobs.

First some additional notation should be introduced. $\Delta_\lambda^x$ is the $x$-th percentile of the relative error obtained from the cumulative distribution function (CDF) of the relative error obtained from the historical data, such as the one presented on Figure 5. Also, let $\hat{\lambda}$ be the value predicted using the forecasting tool. Then it is legitimate to say that:

[2]http://www.neo.ne.gov/statshtml/115.htm.

$$Pr(\lambda < (1 + \Delta_\lambda^x)\hat{\lambda}) = x, \qquad (9)$$

where $x$ is some probability. For example, for the Wikipedia traces the 90% percentile produced by Winter's method is 0.09 [tbl. I] and let's say the next $\lambda$ produced by the same forecaster is $12 \times 10^6$. Consequently the probability that the actual $\lambda$ for the next configuration interval being smaller then $13.08 \times 10^6$ is 90%.

Next, we contrast the behavior of the adaptive policy when used with three different arrivals rates adjusted to the 90-th, 95-th and 99-th percentile respectively. The three modifications are contrasted against the static allocation policy (*i.e.*, all servers are switched on) and the adaptive policy algorithm which does not perform any overprovisioning.

As it can be seen from Figure 6 all adaptive search modifications provide better results in terms of revenue and significantly outperform the static allocation policy. Moreover, the results exhibited by the modification of the 'Adaptive' policy are extremely close, even though the 'Adaptive' algorithm used with 95-th percentile shows slightly better results.

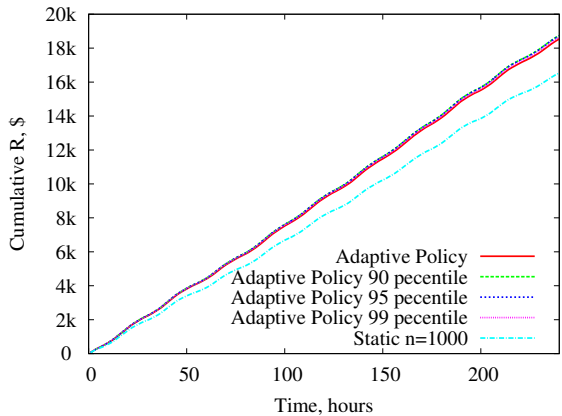

Fig. 6. Cumulative revenue.

As for the amount of consumed energy, as expected the 'Adaptive' policy without any overprovisioning is the one achieving the best result, see Figure 7, with the worst results being produced by the static allocation policy. The adaptive policy used with 99-th percentile is 'overcautious' which adversely reflects on the energy consumption. However, it terms of lost jobs the difference between the static allocation policy and the adaptive policy used with 99-th percentile are almost indistinguishable. At the same time, the difference between the pure 'Adaptive' policy and the one using 90-th percentile is almost of one order of magnitude, see Figure 8.

Comparing the results in all three categories makes us conclude that employing the 'Adaptive' policy with 90-th or 95-th percentile represents the rational trade-off between the optimality of the generated revenue, the amount of consumed energy and the number of lost jobs.

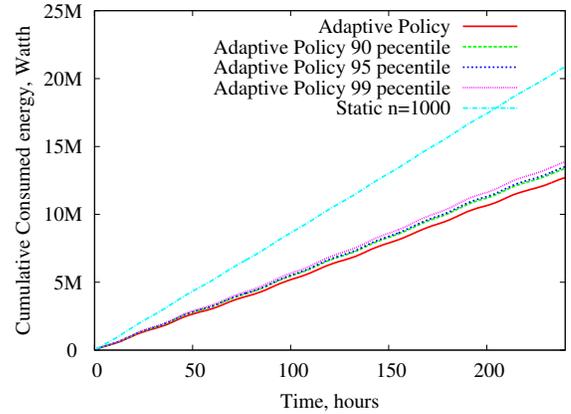

Fig. 7. Cumulative consumed energy.

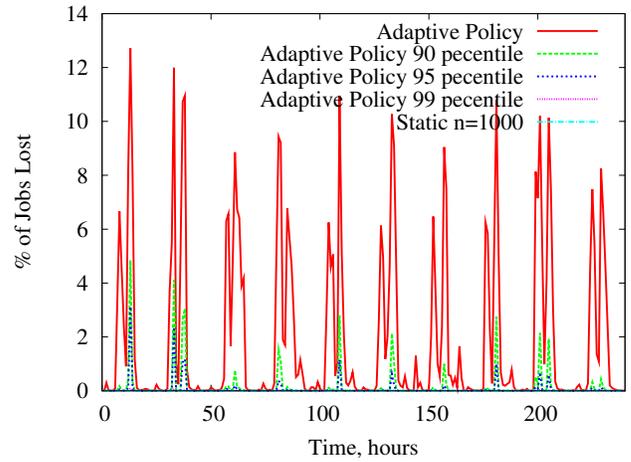

Fig. 8. Relative number of jobs lost.

### C. QED policy

A possible alternative to the 'Adaptive' algorithm is the QED policy which, even though not optimal, is faster and directly addresses the case when the information about the arrival rate is not precise. The QED policy has a parameter, $\alpha$, which needs to be estimated in order to optimize its efficiency. Unfortunately, the complexity of the revenue function does not allow a simple analytical solution and thus we resort to an empirical evaluation of different values of $\alpha$, *e.g.*, the probability that all servers are busy, in order to find the one producing the best results. In the next set of experiments we consider $\alpha$ taking the values of 0.4, 0.2, 0.1, 0.05, 0.025 and 0.01. Once again the approach is contrasted against the static allocation policy (and to the 'Adaptive' algorithm by using the figures described in the previous subsection). The first observation is that the value of $\alpha$ does not have a significant impact on the achieved revenue, as all QED versions exhibit results which are markedly better then that produced by the static allocation policy, see Figure 9. On the other hand $\alpha$ has a strong effect on the number of lost jobs, see Figure 10. Since the traffic is rather periodic, the policies exhibit the same behaviour with the period of 24 hours; hence, in order to improve the readability of the chart we show only the first 48 hours. While QED with $\alpha$ greater then 0.20 shows unacceptable results, the versions with $\alpha$ lower then 0.025 perform very well. In terms of energy consumption, all versions of QED behave in a very similar manner, and significantly outperform the static allocation policy, see Figure 11. Also, it is easy to observe that, even though the QED algorithm is much easier than the 'Adaptive' policy, they achieve about the same level of revenues and consume almost the same amount of energy. Hence, we can conclude that the QED policy with $\alpha < 0.025$ represents a good candidate for the deployment in a real system.

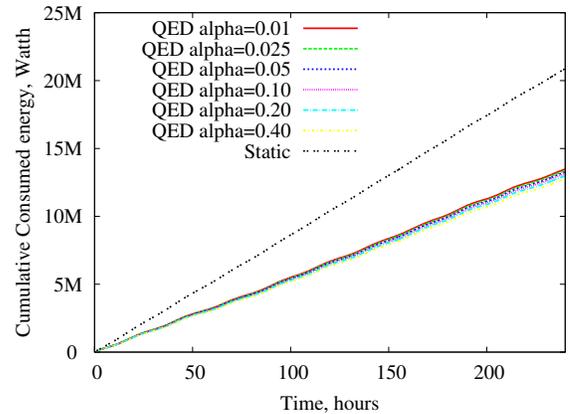

Fig. 11. Cumulative consumed energy.

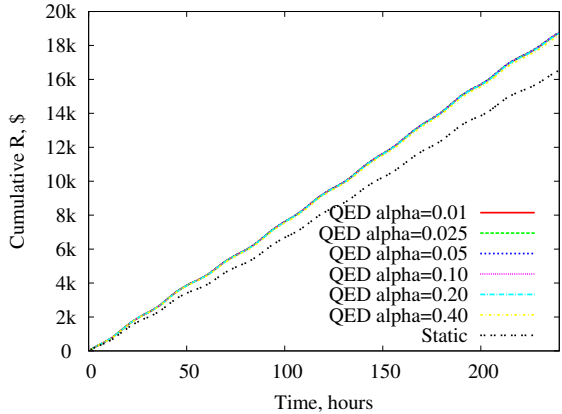

Fig. 9. Cumulative revenue.

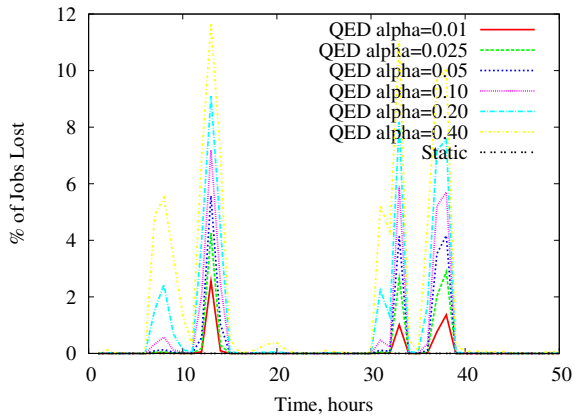

Fig. 10. Relative number of jobs lost.

## V. RELATED WORK

As large scale distributed systems gather and share more and more computing nodes and storage resources their energy consumption rises very quickly. On the other hand, in order to make a fully efficient use of computing and communication systems and reduce their environmental and social impact, green computing and communication research has increasingly become a hot topic of interest for both the computer research community and industry. Unfortunately, while the problem of energy efficiency in mobile devices and laptops has already attracted a lot of attention, the problem of energy efficiency in data centers remains under researched. All the efforts in this area are either $(i)$ intensive, *i.e.*, optimizing power consumption of a server, for example by dynamically changing the CPU voltage/frequency, or $(ii)$ extensive, *i.e.*, minimizing the power consumption of a server pool, for example by dynamically switching some servers off when they are not needed,

Most of the intensive approaches have tried to minimize the power consumption when the number of servers is fixed. While Google engineers have called for systems designers to develop servers that consume energy in proportion to the amount of computing work they perform [13] and Microsoft engineers have been working on better power management on the operating system layer [14], the reality is that servers consume as much as 65% of their peak power when idle [15]. Hence, Elnozahy et al. [16] investigated the potential benefits of scaling down the CPU voltage/frequency according to the offered load, finding that savings can be as big as 20–29%, while Wierman et al. [17] and Horvath et el. [18] considered changing the voltage of servers' CPU in order to minimize the power consumption while trying to meet certain predefined deadlines.

Most of the research conducted using extensive methodologies considered scenarios where the number of running servers can be controlled at runtime. Thus, the idea is to power down some servers whenever that can be justified by demand conditions. For example, Chase et al. [1] presented an architecture for resource management of server farms. There the goal is to reduce energy consumption, while the SLAs are assumed to be flexible (*i.e.*, service degradation is a viable option). The most closely related work can perhaps be found in [19] and [20]. [19] presents a queuing model for controlling the energy consumption of service provisioning systems subject to Service Level Agreements (SLAs). However, while Chen et al. take into account the cost for smaller mean time between failures (MTBF) when powering up/down some servers, the cost function they propose does not consider the time and energy wasted during state changes, nor the cost for failing to meet the promised quality requirements. Hence, the taken decisions could be either too performance oriented or too energy-efficiency oriented. [20], instead, discusses a problem similar to that we attack in this paper. However, in that paper the authors assume that clients have no patience, while they do not consider the fact that servers consume energy without producing any revenue during system reconfigurations.

Finally, since running too many servers increases the electricity consumption while having too few servers switched on requires running those servers' CPUs at higher frequencies, some hybrid approaches have been proposed. For example, [16] attempts to find a rational trade off between the number of servers switched on and the voltage/frequency of the CPU on each server.

## VI. FUTURE WORK

In this paper we have evaluated the performance of the proposed policies via simulation. Next, we plan to expand this work by implementing a prototype mimicking Wikipedia setup with real servers.

Our model is currently designed to optimize the energy consumption of a single tiered system. However, a significant percentage of systems is implemented as two or three tiered, where one tier can be used for realizing caching, the second hosts application logic and the third is responsible for accommodating database servers. Thus, we plan to deal with scenarios where more than one tier is used. Also, we plan to assess the performance of the proposed policies under different workloads (*e.g.*, compute intensive, communication intensive and storage intensive).

Other issues we plan to deal with include electricity charges and PUE factor. Currently we assume that the service provider is charged a flat fee for electricity consumption. However, when it comes to purchasing electricity in large amounts the price fluctuates depending on the time of the day, day of the week, *etc*. Factoring in dynamic prices for electricity constitutes another direction of the future development of this work. Finally, depending on the size and type of the data center different cooling systems can be employed. Elaborating the model in order to incorporate the cooling costs depending on the type of the cooling system is another possible future direction.

## VII. CONCLUSIONS

In this paper we have introduced and evaluated some easily implementable policies for dynamically adaptable Internet services.

Under some simplifying assumptions, the numerical algorithm implemented by the 'Adaptive' policy can find the best trade off between consumed power and delivered service quality, while the QED heuristic we have introduced performs well. A special focus has been given to issues related to the quality of the workload estimation (*i.e.*, predicting the future arrival rate). The enhancement adressing the latter problem were suggested and evaluated using Wikipedia traces, showing significant improvements over the original policies which assume that the parameters are known and accurate.

## ACKNOWLEDGEMENTS

The authors would like to thank the EU Cost Action IC0804 (Energy Efficiency in Large Scale Distributed Systems), and the EUREKA Project 4989 (SITIO).